\documentclass{JHEP3}
\usepackage{graphics,amsmath,mathrsfs}
\usepackage{epsfig}

\title{A Comparison of two different jet algorithms for the top mass reconstruction at the LHC}
\author{Michael H. Seymour\\
School of Physics \& Astronomy, University of Manchester, U.K.; and\\
Theoretical Physics Group, CERN, CH-1211 Geneva 23, Switzerland.}

\author{Christopher Tevlin\\
School of Physics \& Astronomy, University of Manchester, U.K.}

\abstract{We compare the abilities of the cluster-type jet algorithm, KtJet, and a mid-point iterating cone algorithm to reconstruct the top mass at the LHC. We discuss the information contained in the merging scales of cluster-type algorithms, and how this can be used in experimental analyses, as well as the different sources of systematic errors for the two algorithms. We find that the sources of systematic error are different for the two algorithms, which may help to better constrain the systematic error on the top mass at the LHC.}

\preprint{CERN-PH-TH/2006-075\\ MAN/HEP/2006/16}

\begin{document}
\noindent
\section{Introduction}

In the Standard Model (SM), the theory that describes the strong interaction is Quantum Chromodynamics (QCD). This is a gauge theory which describes the interactions of coloured partons, called quarks and gluons. This has been very successful in describing a large range of experimental data, and since it has a negative beta function it naturally explains the concepts of asymptotic freedom, and confinement. At high energy colliders, one can exploit the concept of asymptotic freedom, and use perturbation theory to calculate the cross sections for certain partonic final states. However these coloured partons do not propagate over macroscopic distances, and will always become confined into colour singlet states, called hadrons. In fact one observes well collimated jets of hadrons, whose momenta closely reflect the momenta of the final state partonic configuration, and it is these that are observed in such experiments. In order to test QCD precisely, one needs an unambiguous definition of a jet, or a `jet algorithm', for putting hadrons together in groups which may be identified with the original final state partons. Clearly since colour is conserved no unique mapping exists, so different choices have been made.

One type of jet algorithm is a Cone Algorithm~\cite{Sterman:1977wj}, which, in its simplest form, combines all hadrons in a cone of fixed angular extent, and identifies these with one of the initial partons. There are two main problems with such an algorithm. Firstly, this is not a one-to-one mapping, so one hadron may be assigned to more than one jet. Some prescription for removing overlaps between jets is therefore required. Secondly, in its simplest form the algorithm is not infrared/collinear safe. Before discussing the issue of infrared safety\footnote{It is common to use the term infrared safety to mean both infrared and collinear safety, this will always be implied in this paper.}, we note that the cone algorithm may be extended~\cite{Abe:1991ui,Ellis:1992qq} so that neither of these problems persist. Such an algorithm has been used in this analysis, and will be defined in Section~\ref{jetalg}.

The definition of infrared safety is:

\begin{flushleft}
\it{An observable is infrared safe if, for any n-parton configuration, adding an infinitely soft parton does not affect the observable at all.}
\end{flushleft}

Similarly the definition of collinear safety is:

\begin{flushleft}
\it{An observable is collinear safe if, for any n-parton configuration, replacing any massless parton by an exactly collinear pair of massless partons does not affect the observable at all.}
\end{flushleft}

It is a well known aspect of gauge theories (or more generally any theory which contains massless fields), that at next to leading order (NLO), one encounters infrared and collinear singularities in expressions for individual Feynman diagrams. It can easily be show that for any IR safe observable as defined above, all such singularities cancel yielding finite predictions. Another way to view this is that the `jet structure' of an event should be insensitive to arbitrarily soft or collinear emissions.

Another type of jet algorithm is a Clustering Algorithm, examples of which are that used by the JADE collaboration~\cite{Bartel:1986ua,Bethke:1988zc} and, more recently, the QCD motivated $k_{\perp}$ algorithm~\cite{Catani:1991hj}, which has been extended to hadron-hadron collisions~\cite{Catani:1993hr,Ellis:1993tq}. Although the specific algorithm used in this analysis will be defined in Section~\ref{jetalg}, here we shall outline the general idea for hadron colliders. One defines a closeness measure $d_{iB}$ between each object $i$ and the beam direction, and similarly a closeness measure $d_{ij}$ between each pair of objects $i$ and $j$. The smallest object in the set $\{d_{iB},d_{ij}\}$ is found; if it is a $d_{iB}$ then object $i$ is considered to belong to the beam in some way, and if it is a $d_{ij}$ then objects $i$ and $j$ should be merged in some way. Such an algorithm is a one-to-one mapping and so each object is assigned to one and only one jet. The IR safety of this algorithm may be ensured by choosing the closeness measures carefully, for example using the transverse momentum of the particle for $d_{iB}$, and the transverse momentum of the softer object with respect to the direction of the harder object for $d_{ij}$, as will be discussed in Section~\ref{jetalg}.

In an effort to motivate the comparison in this analysis, and in particular motivate clustering algorithms in general we point out the following possible theoretical advantages. Firstly, our understanding of the perturbative splittings which a system of partons undergoes before hadronisation suggests that the cluster algorithm gives a closer reflection of the underlying partonic dynamics. One could also argue that since the cluster algorithm outlined above is manifestly IR safe and assigns each hadron to one jet, it may be theoretically favoured. However, as mentioned above these problems can be overcome in the case of a cone algorithm at the expense of a simple algorithm, so this is perhaps a matter of aesthetics. Every jet algorithm contains at least one parameter that can be tuned to give a configuration of jets that best describes the partonic configuration of one's signal (e.g. the cone radius in cone algorithms, or $y_{cut}$ in cluster algorithms). These parameters tend to enter into cross sections logarithmically, and by tuning these one can cause these terms to become large. In this case the convergence of the perturbative series can be spoilt, and one needs to perform a resummation of these large terms to all orders. This is possible in the case of the $k_{\perp}$ cluster algorithm, but not in the case of the cone algorithm.

It is worth mentioning at this stage that the cluster algorithm contains much more information about an event than simply the number of jets and their 4-momenta. One has access to the different merging scales within the event which are perturbative quantities that can be used in experimental analyses. We shall give a more detailed account of these, and how they are exploited in this study in Section~\ref{jetalg}.

When comparing the performance of two jet algorithms experimentally, there is an arbitrariness; what makes one algorithm better than another? This is (slightly) less ambiguous when reconstructing a massive particle. One knows the mass, momentum etc. of the decaying particle, and can compare this to that reconstructed by the jet algorithm. This makes the top mass measurement an ideal place to study the advantages/disadvantages of one algorithm over another. However, it must be emphasized that the top mass measurement to a high precision will be one of the very important measurements made at the LHC. Such a measurement would constrain the mass of the Standard Model Higgs boson, as well as many Beyond the Standard Model theories. We present here an update of the study published in~\cite{Seymour:1993mx}. The aim of this study is to explore the possibility of minimising the error on this measurement by using a cluster algorithm. In Section~\ref{syst} we shall discuss the different contributions to the systematic error on the top mass measurement for each of the two jet algorithms. 
\section{Jet Algorithms}
\label{jetalg}

In this section we shall define the jet algorithms that have been used in this study. In sections~\ref{cone} and~\ref{ktjet} we shall outline the (infrared safe) mid-point iterating cone algorithm and the cluster algorithm implemented in the KtJet~\cite{Butterworth:2002xg} package respectively. In Section~\ref{mymode}, we shall outline precisely how KtJet was used in this study. 

\subsection{Mid-point Iterating Cone Algorithm}
\label{cone}

There are very many variations of the cone algorithm, and we present an outline of the cone algorithm used in this study. This is a C++ implementation of a cone algorithm~\cite{will}, based on the Fortran algorithm PxCone~\cite{pxcone}. The definition is as follows:
\begin{enumerate}
\item Call every final state object a `seed direction', ($\eta_{S},\phi_{S}$).
\item For each seed direction, calculate the jet momentum,
\begin{equation}
E_{TJ}=\sum_{i\epsilon J}E_{Ti}\ ,
\label{cone1}
\end{equation}
\begin{equation}
\eta_{J}=\frac{1}{E_{TJ}}\sum_{i\epsilon J}E_{Ti}\eta_{i}\ ,
\end{equation}
\begin{equation}
\phi_{J}=\frac{1}{E_{TJ}}\sum_{i\epsilon J}E_{Ti}\phi_{i}\ ,
\label{cone3}
\end{equation}
\begin{equation}
{\rm where}\  i\epsilon J\Leftrightarrow (\eta_{i}-\eta_{S})^{2}+(\phi_{i}-\phi_{S})^{2}<R^{2}\ .
\end{equation}
\item If the jet and seed directions are not equal, then define the new seed direction to be the jet direction, and go to step 2. 
\item If the jet is not the same as any already found, add it to a list of `protojets'.
\item Consider the mid-point of every pair of protojets which are separated by 1-2 cone radii found in steps 1 to 4 as a seed direction, and repeat steps 2 to 4.
\item Delete all protojets that have a transverse energy less than 6 GeV.
\item Delete all protojets that have more than 75\%, of their transverse energy contained within higher transverse energy protojets.
\item Assign all particles in more than one protojet to the one whose centre is nearest in ($\eta,\phi$), recalculating the jet momenta using~(\ref{cone1})-(\ref{cone3}). All protojets remaining at this stage are called jets.
\end{enumerate}

Step 5 ensures the infrared safety of the algorithm, while Step 8 ensures that it uniquely assigns each object to at most one jet. 

\subsection[The $k_{\perp}$ Algorithm]{The \boldmath$k_{\perp}$ Algorithm}
\label{ktjet}

The $k_{\perp}$ Algorithm is very flexible, and allows the user several choices when defining jets. Here we shall give an outline of the $k_{\perp}$ Algorithm as it was implemented in this study (the exclusive mode) and for a more general definition of the algorithm we refer the reader to~\cite{Butterworth:2002xg}. The algorithm is as follows:
\begin{enumerate}
\item For every final state object $h_{i}$ and for each pair of objects $h_{i}$ and $h_{j}$, compute the variables $d_{iB}$ and $d_{ij}$ given by:
\begin{equation}
d_{iB}=p^{2}_{ti}\ , 
\end{equation}
\begin{equation}
d_{ij}={\rm min}(p^{2}_{ti},p^{2}_{tj})R^{2}_{ij}\ ,
\end{equation}
where
\begin{equation}
R^{2}_{ij}=(\eta_{i}-\eta_{j})^{2}+(\phi_{i}-\phi_{j})^{2}\ .
\end{equation}
\item Find $d_{min}$, the smallest element of the set $\{d_{iB},d_{ij}\}$. If a $d_{ij}$ is the smallest, then the two objects $h_{i}$ and $h_{j}$ are combined into a single object with momentum $p_{(ij)}$ such that:
\begin{equation}
p_{t(ij)}=p_{ti}+p_{tj}\ ,
\end{equation}
\begin{equation}
\eta_{ij}=\frac{p_{ti}\eta_{i}+p_{tj}\eta_{j}}{p_{t(ij)}}\ ,
\end{equation}
\begin{equation}
\phi_{ij}=\frac{p_{ti}\phi_{i}+p_{tj}\phi_{j}}{p_{t(ij)}}\ .
\end{equation}
If a $d_{iB}$ is the smallest, then the object $h_{i}$ is included in a `beam jet' and removed from the list.
\item If the number of jets is equal to the stopping multiplicity, N, then the algorithm is complete, and all of the remaining objects will be classified as jets. If not go back to (1).
\end{enumerate}

Note that one can stop the merging when $d_{min}>d_{cut}$, where $d_{cut}$ is a stopping parameter with dimensions of energy squared. The choice of jet multiplicity to use will be discussed in Section~\ref{mymode}.

Finally, we point out that when running KtJet in the exclusive mode, the algorithm attempts to factorise the hard scatter from the soft underlying event. When running KtJet in this mode, it is useful to check how well the algorithm does this, i.e.~will the jets be a good reflection of the partons from the hard scatter. This will be discussed briefly in Section~\ref{toy}.

\subsection{Final Jet Multiplicities}
\label{mymode}

In the Standard Model, top quarks decay almost exclusively into a $W$ boson and a $b$ quark. The $W$ boson is approximately on-shell and may decay hadronically into a $q\bar{q}^{\prime}$ pair, typically resulting in two jets; or leptonically into a charged lepton and a neutrino. In this study our signal process is the so called `lepton plus jets' channel, which we define to be inclusive $t\bar{t}$ production with one of the $W$ bosons decaying leptonically, the other hadronically.

\begin{figure}[ht]
\begin{center}
\includegraphics[height=4cm]{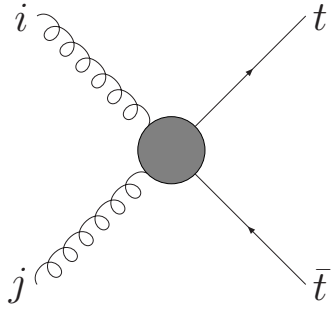}
\caption{Schematic representation of the LO Feynman diagrams for $t\bar{t}$ production.}
\label{ttbar-lo}
\end{center}
\end{figure}

The phenomenological signature will be two $b$-jets, two light jets, one hard, isolated lepton, and some missing $p_{t}$. So as a first guess, one might try a final jet multiplicity of four. However, the presence of additional hard partons in the final state, will give rise to additional high $p_{t}$ jets. The leading order Feynman diagrams for $t\bar{t}$ production are shown schematically in Figure~\ref{ttbar-lo}. At NLO there are both real and virtual contributions to this process, shown schematically in Figure~\ref{ttbar-nlo}. As a result of the real contributions, one would expect to observe a fraction of events with an additional hard jet from the additional gluon. Thus one would expect the lepton plus jets channel to have a contribution from some events with one lepton and five hard, well separated jets. 
\begin{center}
\end{center}
\EPSFIGURE[ht]{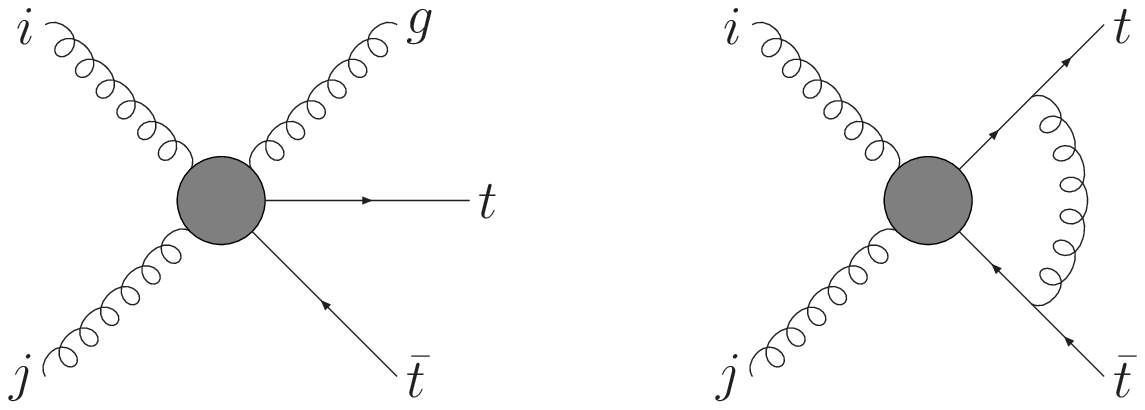}{Schematic representation of real and virtual NLO Feynman diagrams for $t\bar{t}$ production.\label{ttbar-nlo}}

\begin{figure}[ht]
\begin{center}
\includegraphics[height=7.5cm]{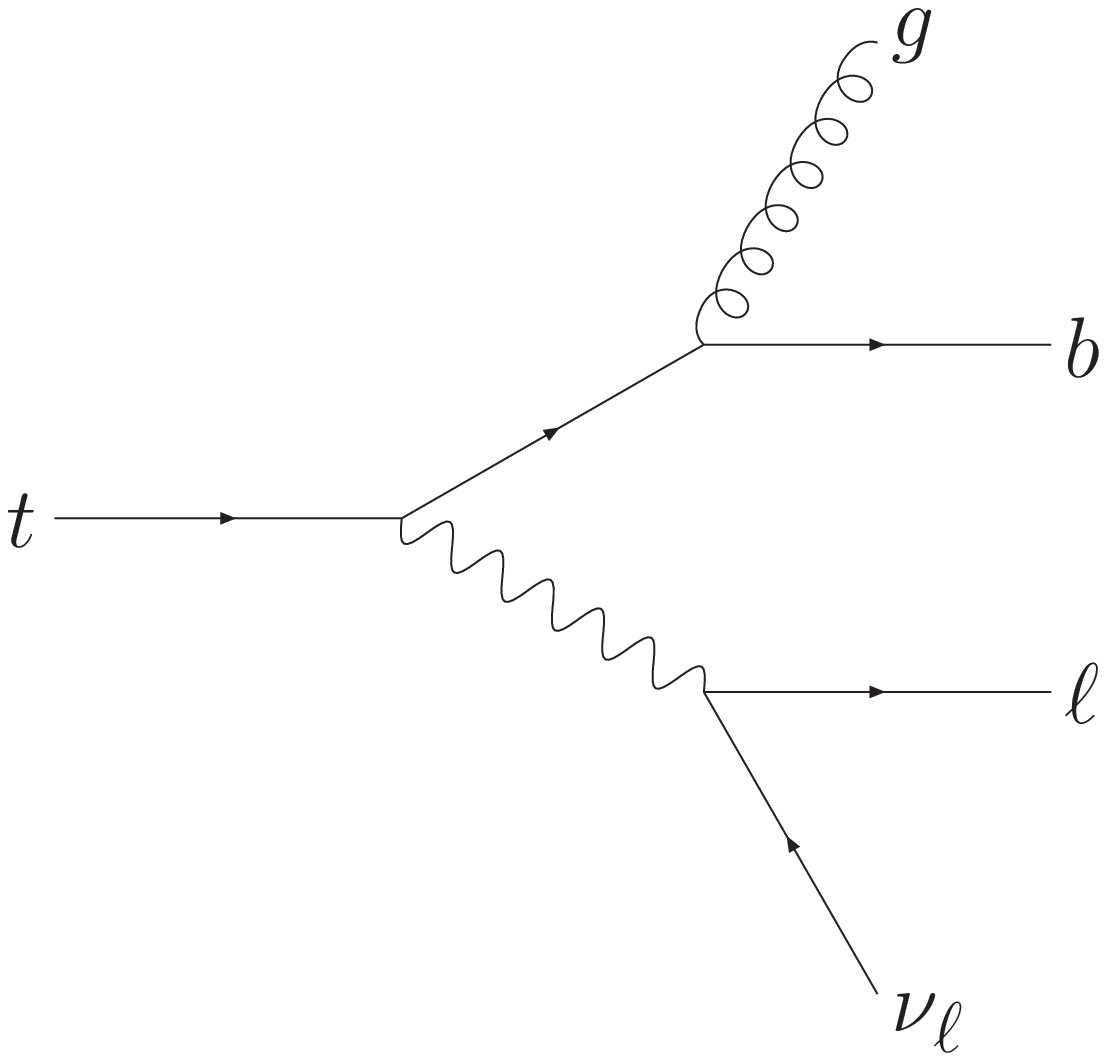}
\caption{A higher order correction to top decay.}
\label{topdecay}
\end{center}
\end{figure}

The only other source of additional high $p_{t}$ jets (excluding the possibility of a hard jet from secondary scatters) is final state radiation from the decay products of the top quarks. Thus in some fraction of events, one of the top quarks could decay into either four jets, or a lepton-neutrino pair and two jets. If the leptonically-decaying top decays as in Figure~\ref{topdecay}, then one would again expect to observe one lepton and five jets.

If the final state really contains five hard well separated jets, then by clustering to four jets, one will either merge two of these jets together or remove the softest jet. So the hadronically-decaying $W$ boson may not be resolved as two jets, or one of the three jets from the top decay may be removed. In such events, one needs to choose a higher final jet multiplicity. However, in such events, one would expect the scales at which $i+1$ objects are merged into $i$ objects, $d_{i+1,i}^{1/2}$, to be quite different, especially the scale $d_{5,4}^{1/2}$, which one would expect to be much larger for 5 jet events. These merging scales are safe perturbative quantities, which are all accessible in the KtJet package.

In order to decide when to stop merging at a multiplicity of four, and when to stop sooner at, say five, we generated two samples of $t\bar{t}+0jets$ and $t\bar{t}+1jet$ events, with the Monte Carlo event generator ALPGEN~\cite{Mangano:2002ea}, imposing the MLM matching prescription~\cite{mlm}. The KtJet merging scales for the two samples are shown in Figure~\ref{toy1}.
\EPSFIGURE[ht]{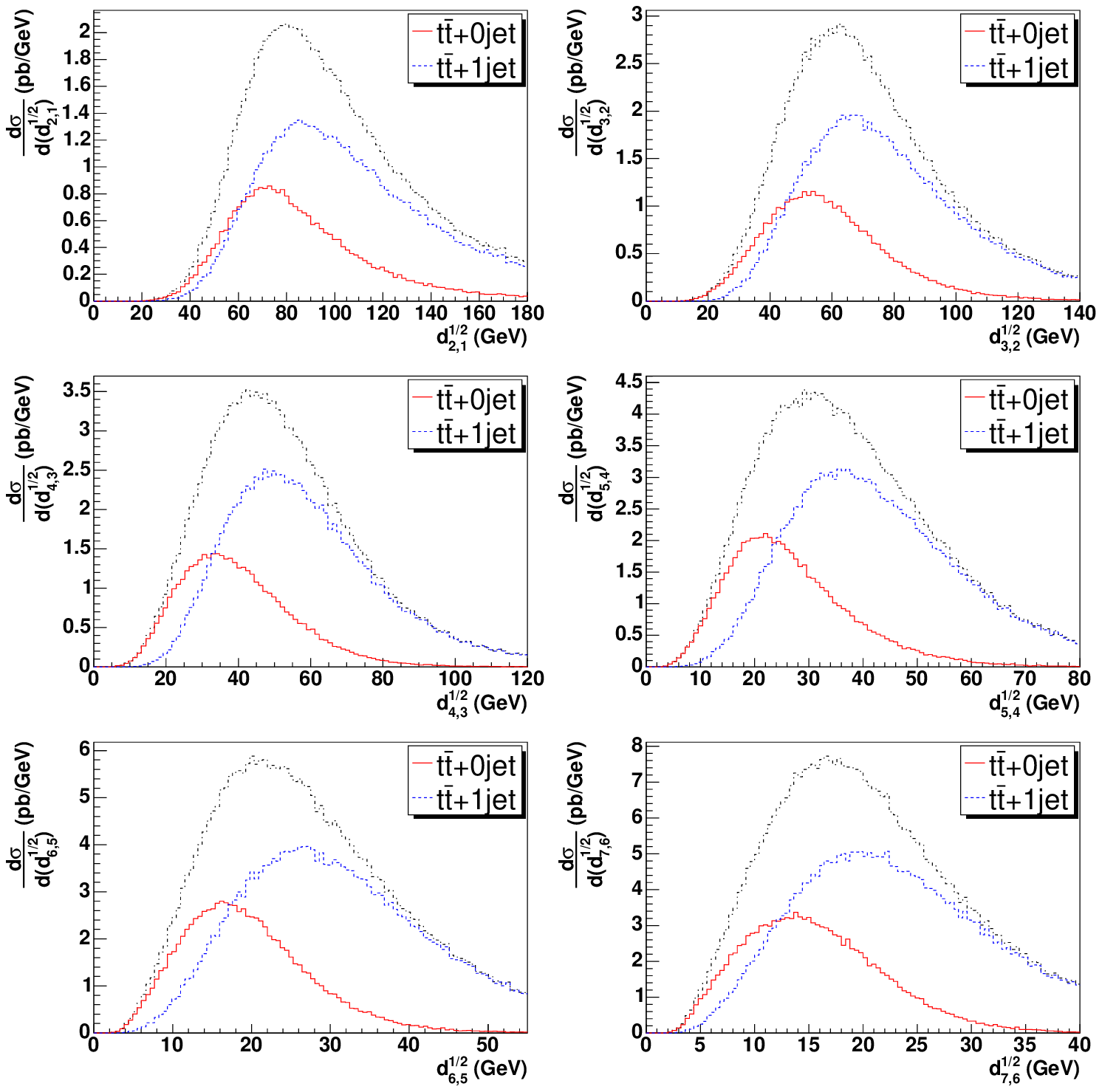}{The highest six merging scales from KtJet for samples of (a) $t\bar{t}$~(red, solid), (b) $t\bar{t}+1jet$~(blue, dashed) and (c) the total~(black, dot-dashed); generated with ALPGEN.\label{toy1}}
These do indeed seem to be promising variables to use to decide on a final jet multiplicity, however, they are correlated. This means that making several independent 1-dimensional cuts could be difficult to optimise. For this reason, we form the Fisher discriminant, $\mathscr{F}$, which is the linear combination of these variables that maximises the separation between the two samples. For these six variables, $\mathscr{F}$ is given by
\begin{equation}
\mathscr{F}=0.053\frac{d^{1/2}_{2,1}}{78}+0.122\frac{d^{1/2}_{3,2}}{63}+0.185\frac{d^{1/2}_{4,3}}{45}+0.414\frac{d^{1/2}_{5,4}}{28}+0.404\frac{d^{1/2}_{6,5}}{20}+0.099\frac{d^{1/2}_{7,6}}{16}\ ,
\end{equation}
where we have divided the merging scales by the approximate peak value, so that the relative importance of the different merging scales in separating the two samples can be seen. The scales $d^{1/2}_{5,4}$ and $d^{1/2}_{6,5}$ are the most important scales. This Fisher Discriminant is shown in Figure~\ref{fisher} for the two samples. It is this that we shall use to decide on the final jet multiplicity. Precisely where to cut, and how to optimise such a cut, will be discussed in Section~\ref{analysis}.
\EPSFIGURE[ht]{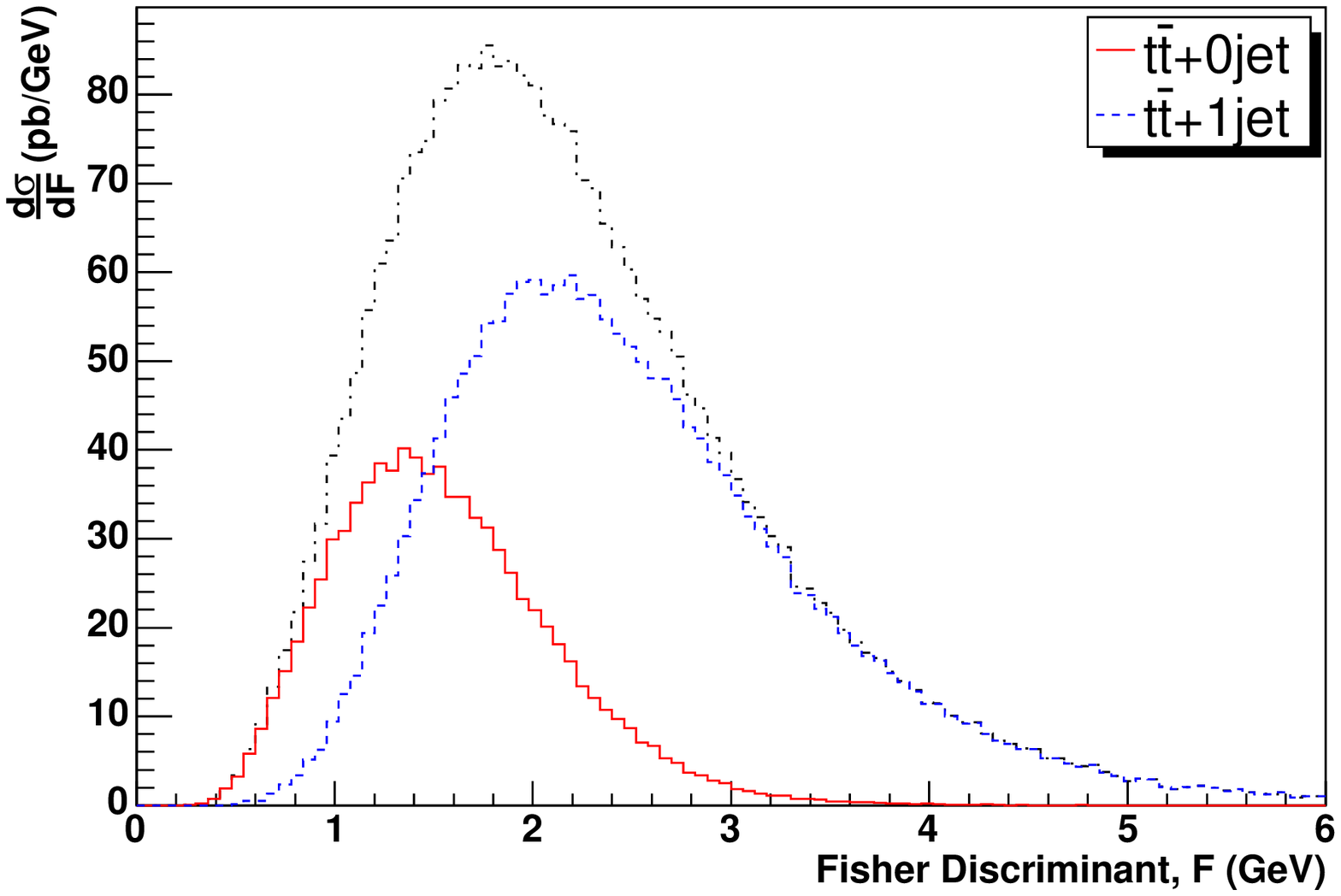,height=8cm}{The Fisher Discriminant formed from a linear combination of the variables in Figure~\ref{toy1}.\label{fisher}}
Finally, we note that using this procedure to decide when to stop merging in KtJet is analogous to deciding what cone radius to use in a cone algorithm. One could generate a sample of events, and calculate the typical boosts of certain particles, and so decide on an optimum cone radius. In Section~\ref{analysis}, we shall find the optimal cone radius for reconstructing tops in this study.
\subsection{MC simulations of jet merging scales}
\label{toy}

As pointed out above, when running KtJet in the exclusive mode, one may wish to ask how well the algorithm has factorized the hard sub-process from the soft underlying event. For a sample of $t\bar{t}$ events, generated by MC@NLO\footnote{For precise details of the sample see Section~\ref{event}.}~\cite{Frixione:2002ik,Frixione:2003ei}, we ran KtJet over the output at various stages of event processing: the partons that came from the hard sub-process, after the parton shower (i.e.~including initial and final state radiation); all partons in the event after the parton shower (i.e.~also including partons from multiple scatters); as well as the final hadron level output. In this section we compare the KtJet merging scales for these three cases, which are shown in Figure~\ref{toy2}. We use this as a measure of how well the rest of the event has been removed by the algorithm.
\EPSFIGURE[ht]{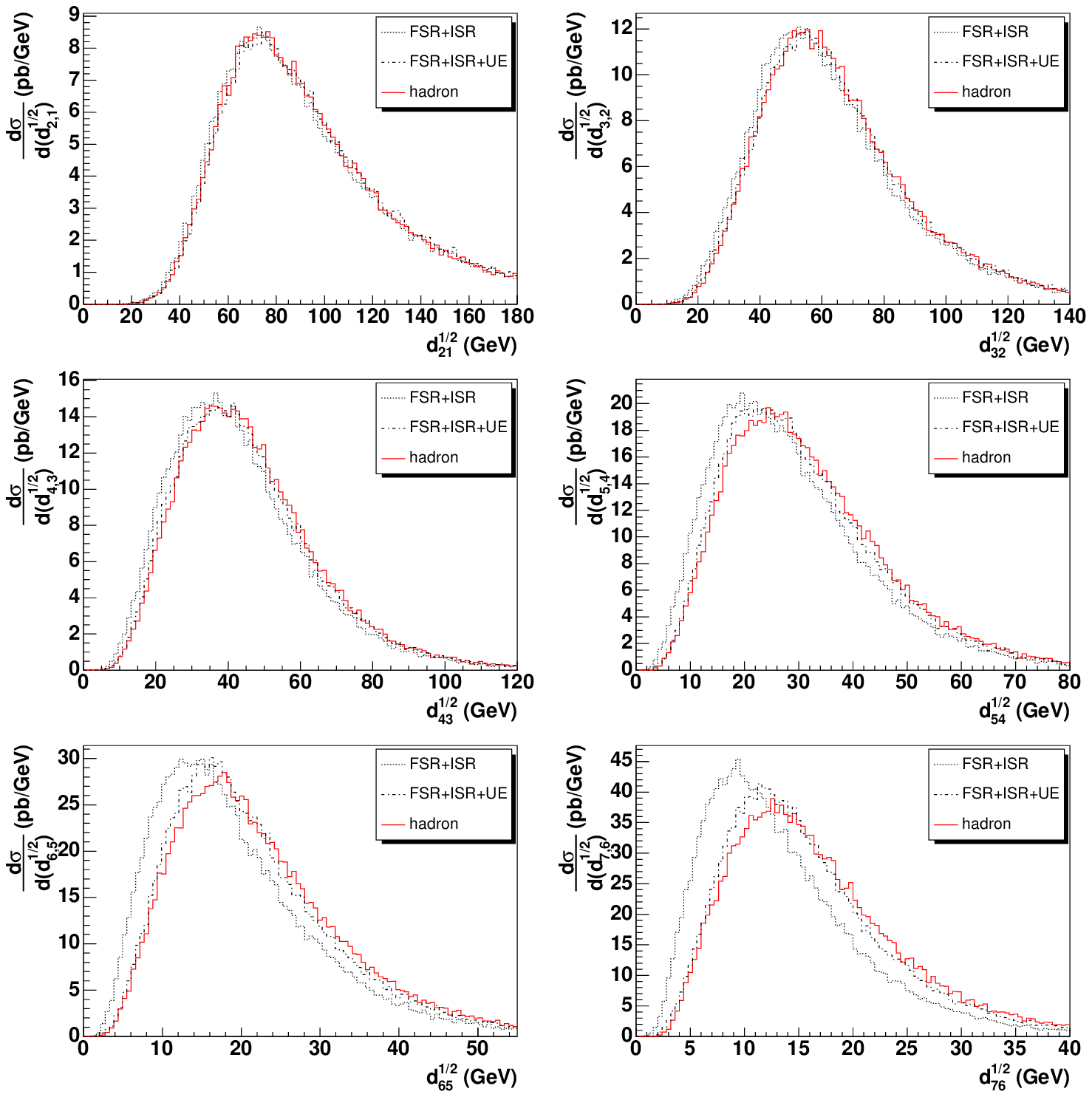}{The highest six merging scales from KtJet with an input of (a) the partons from the hard sub-process after the parton shower~(black, dot), (b) all partons in the event after the parton shower~(black, dot-dash) and (c) the final state hadrons~(red, solid).\label{toy2}}

We compare the highest six merging scales, and we note that there seems to be a reasonable agreement between the three cases. One would expect the hard scatter to give rise to the highest $p_{t}$ objects, and so the highest merging scales should be very similar in the three cases. As one goes down in merging scales, the soft splittings from the parton shower are no longer at a higher scale than the underlying event, and one would therefore expect that the hadron level merging scale, $d_{i+1,i}$, will be higher than the same merging scale at the parton level. This is indeed what we observe - the agreement between hadron level and parton level is very good.

\section{Monte Carlo Event Generation}
\label{event}
As the LHC is not taking data yet, we need to simulate our own. To do this we used the following Monte Carlo event generators. The signal used in this study was a sample of inclusive $t\bar{t}$ events generated by MC@NLO~\cite{Frixione:2002ik,Frixione:2003ei}. This uses a modified subtraction method to match NLO calculations to a leading logarithm (LL) parton shower provided by the Monte Carlo event generator, HERWIG~\cite{herwig}. For simulation of additional (remnant-remnant) scatters, the package JIMMY~\cite{Butterworth:1996zw,jimmy2} was used.

The only background processes considered here are $W+b\bar{b}+jets$ and $W+jets$. We generated samples of $W+b\bar{b}+n$ partons events, for $n=0,1$; and $W+m$ partons events, for $m=0,1,2,3$, with ALPGEN~\cite{Mangano:2002ea}. These samples were then processed by HERWIG for the parton shower and hadronisation and the MLM matching algorithm was imposed to minimise double counting. JIMMY was again used for the underlying event.

\section{Generator Level Analysis}
\label{analysis}

Here we present the main analysis, which closely follows that presented in~\cite{french}. Firstly, we applied the following pre-selection cuts that would be applied to the data to reduce the bulk of the background processes expected at the LHC. The cuts were as follows, there must be:
\begin{enumerate}
\item{At least 20 GeV missing (scalar) transverse momentum,$\not{p_t}$.}
\item{Exactly one hard charged lepton with $p_t>20$ GeV, $|\eta|<2.5$. This must also satisfy the isolation criterion - the amount of energy inside a cone of radius 0.2 centered on the lepton, minus the energy of the lepton must be less than 6 GeV.}
\item{At least 4 jets with $p_t>p_tcut$ GeV, $|\eta|<2.5$, where $p_tcut$ is a variable. We shall show below the dependence of purity and efficiency on $p_tcut$. We also require that exactly two of these are $b$ jets. (We assume a $100\%$ $b$-tagging efficiency).}
\end{enumerate}

The hadronically decaying top is used to determine the top mass, which is the invariant mass of a three jet system: one $b$ jet, and two light jets from the $W$ decay. In Section~\ref{W-alg}, we shall describe the choice of a $W$ candidate, and in Section~\ref{top-alg}, the final top reconstruction. Our signal process is the `lepton plus jets' channel, and the only background considered in this section will be that from other $t\bar{t}$ events, i.e.~combinatorial background.

Finally, in Section~\ref{opt} we shall discuss the optimisation of the two jet algorithms, and compare the performance of each algorithm and how this varies with certain selection cuts. For the cone algorithm we have studied the optimisation of the Cone Radius, whereas for KtJet we have studied the performance as a function of the Fisher Discriminant defined in Section~\ref{mymode}. 
\subsection[$W$ Boson Reconstruction]{\boldmath$W$ Boson Reconstruction}
\label{W-alg}

So far we have required that there be at least two light jets in each event, however when using a cone algorithm, the number of light jets is typically larger than two. Therefore some algorithm must be used to select the two light jets most likely to have come from a $W$ boson, i.e.~the invariant mass of this dijet system must be constrained in some way to the $W$ mass. 

In order to do this, all events with exactly two light jets above the $p_tcut$ were selected\footnote{This selection of events with exactly two light jets was only imposed in order to plot this dijet mass distribution. Events with more than two light jets are still considered as long as a $W$ candidate is found, which has a mass `close' to the peak of this distribution, where close will be defined below.}, and this dijet invariant mass distribution was plotted. Figure~\ref{dijet} shows this distribution for both jet algorithms. 
\EPSFIGURE[ht]{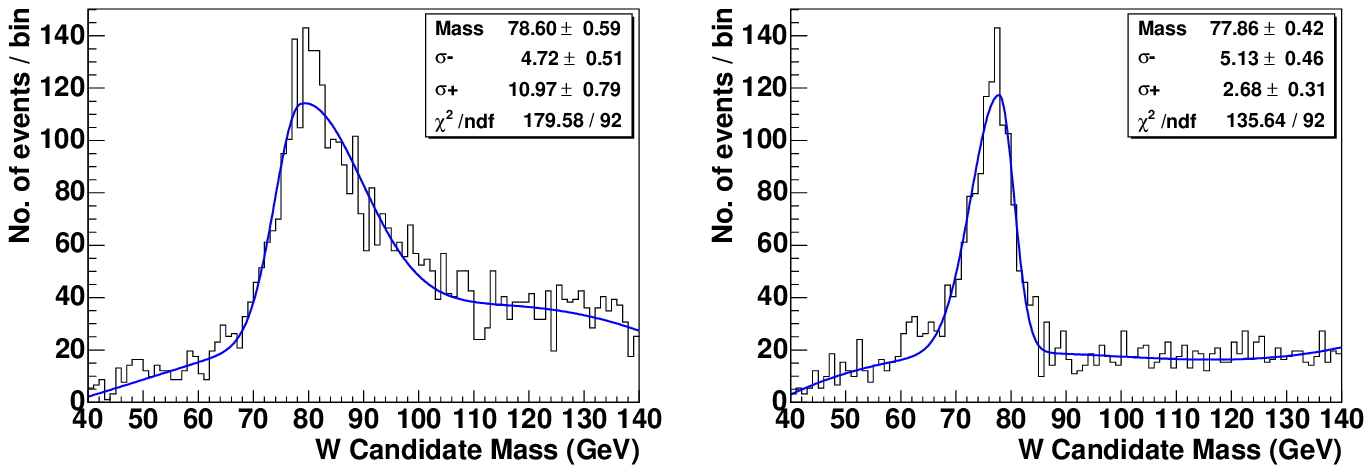}{The dijet invariant mass distribution for events with exactly 2 light jets for (a)~KtJet with a Fisher Cut value of 5.6; (b)~PxCone with a cone Radius of 0.4\label{dijet}}
Since these dijet plots are asymmetric, the sum of two half Gaussians and a polynomial was fitted to the mass distributions, and the best fit values for the peak value, $\langle m_{jj}\rangle$, and the two widths, $\sigma_{jj}^{-}$ and $\sigma_{jj}^{+}$ of these distributions were obtained. In events with more than two light jets, the $W$ candidate is defined as the two light jets that minimise the $\chi^{2}$,
\begin{equation}
\chi^{2}=\frac{(m_{jj}-\langle m_{jj}\rangle)^{2}}{\sigma_{jj}^{2}},
\end{equation}
where
\begin{equation}
\sigma_{jj}=\left\{\begin{array}{ll}
\sigma_{jj}^{-}\ \ \ \ \ &m_{jj}\le \langle m_{jj}\rangle\\\sigma_{jj}^{+}&m_{jj}>\langle m_{jj}\rangle\\
\end{array}\right.\ .
\end{equation}
Only $W$ candidates that lie in a mass window of $[\langle m_{jj}\rangle-q\sigma_{jj}^{-}\ ,\ \langle m_{jj}\rangle+q\sigma_{jj}^{+}]$ are accepted, where $q=1,2,3$ were all considered. Since it was found that increasing $q$ gave little increase in efficiency, and a drop in purity, we shall only present results for $q=1$. Figure~\ref{w-ptcut} shows the purity of the reconstructed $W$ boson as a function of $p_{t}cut$, where purity was defined by comparison to Monte Carlo truth information.
\begin{center}
\EPSFIGURE[ht]{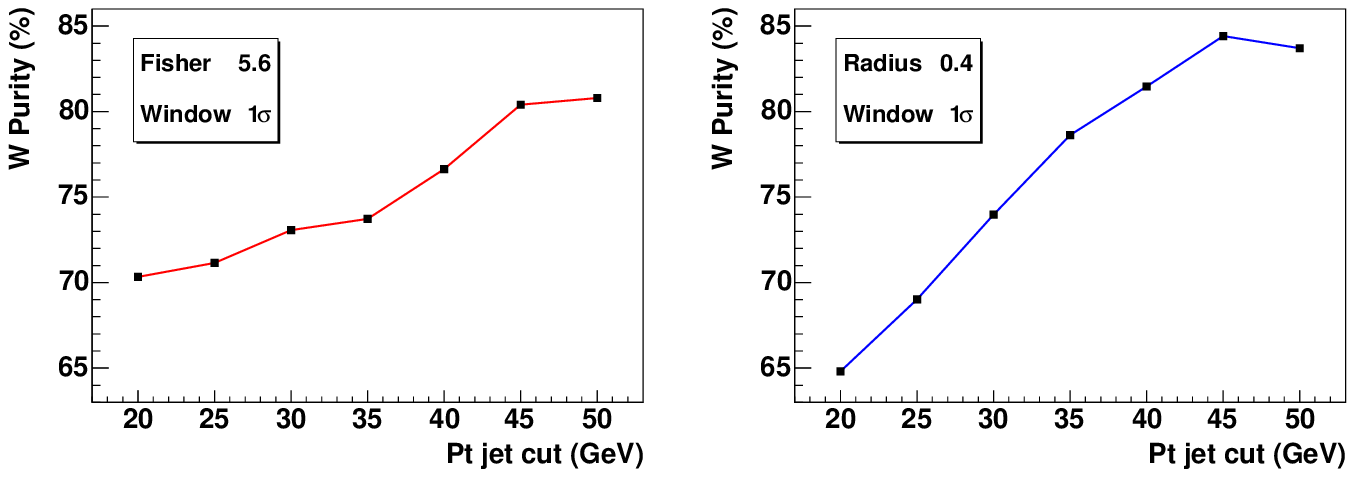}{The purity of the hadronically decaying W boson as a function of $p_{t}cut$ for (a) KtJet with a Fisher Cut value of 5.6; (b) PxCone with a cone Radius of 0.4.\label{w-ptcut}}
\end{center}

\subsection{Top Quark Reconstruction}
\label{top-alg}

Finally to reconstruct the top quark, one must choose one of the two $b$ jets to combine with the $W$ candidate. We choose the $b$ jet which results in the highest $p_T$ top candidate; Figure~\ref{b-purity} shows the purity of the chosen $b$ jet as a function of $p_tcut$.
\EPSFIGURE[ht]{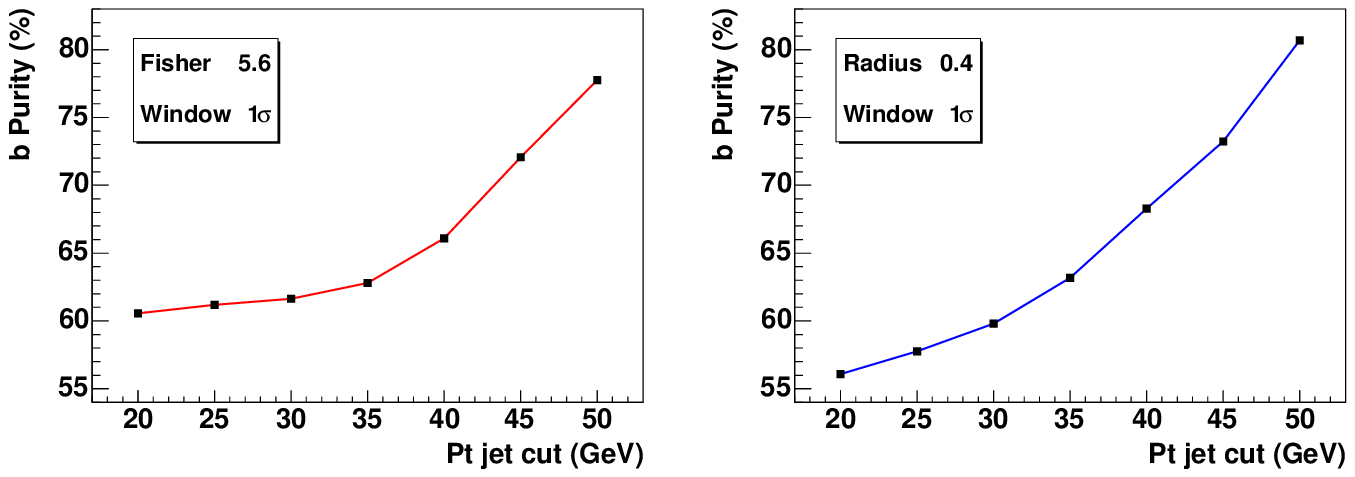}{The purity of the $b$ jet as a function of $p_{t}cut$ for (a) KtJet with a Fisher Cut value of 5.6; (b) PxCone with a cone Radius of 0.4.\label{b-purity}}
The top candidate purity and efficiency are shown in Figures~\ref{top-purity} and~\ref{eff}, and the reconstructed top mass distributions for the two algorithms are shown in Figure~\ref{top-mass}.
The purity of the reconstructed top candidates increases with $p_tcut$, whereas the efficiency decreases. Since the LHC will be a `top factory' with a $t\bar{t}$ event every second, one should be able to make quite harsh selection cuts. In order to select tops with an efficiency of $\approx 1\%$, for both algorithms one should make a $p_tcut$ of 40 - 45GeV. In this case the purity is high for both algorithms.
\EPSFIGURE[ht]{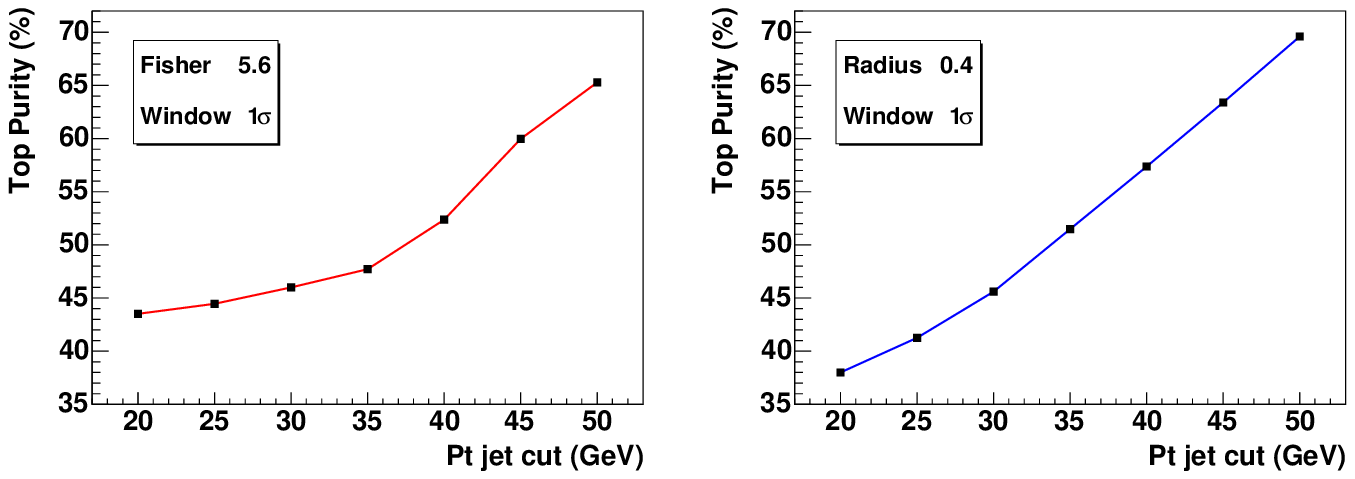}{The purity of the top candidate jet as a function of $p_{t}cut$ for (a) KtJet with a Fisher Cut value of 5.6; (b) PxCone with a cone Radius of 0.4.\label{top-purity}}
\EPSFIGURE[ht]{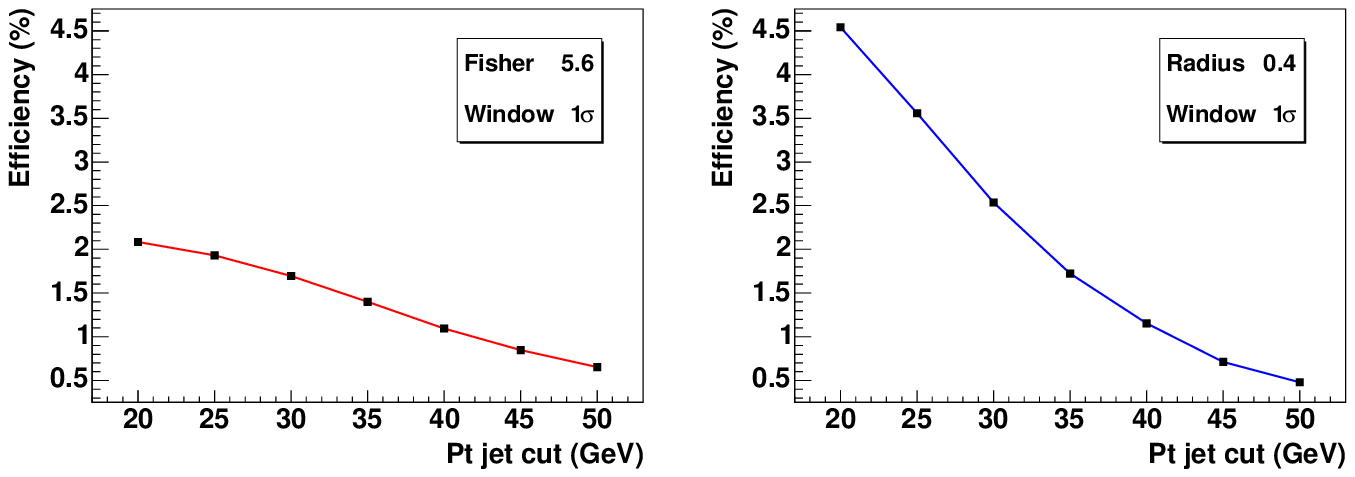}{The efficiency of top reconstruction as a function of $p_{t}cut$ for (a) KtJet with a Fisher Cut value of 5.6; (b) PxCone with a cone Radius of 0.4.\label{eff}}
\EPSFIGURE[ht]{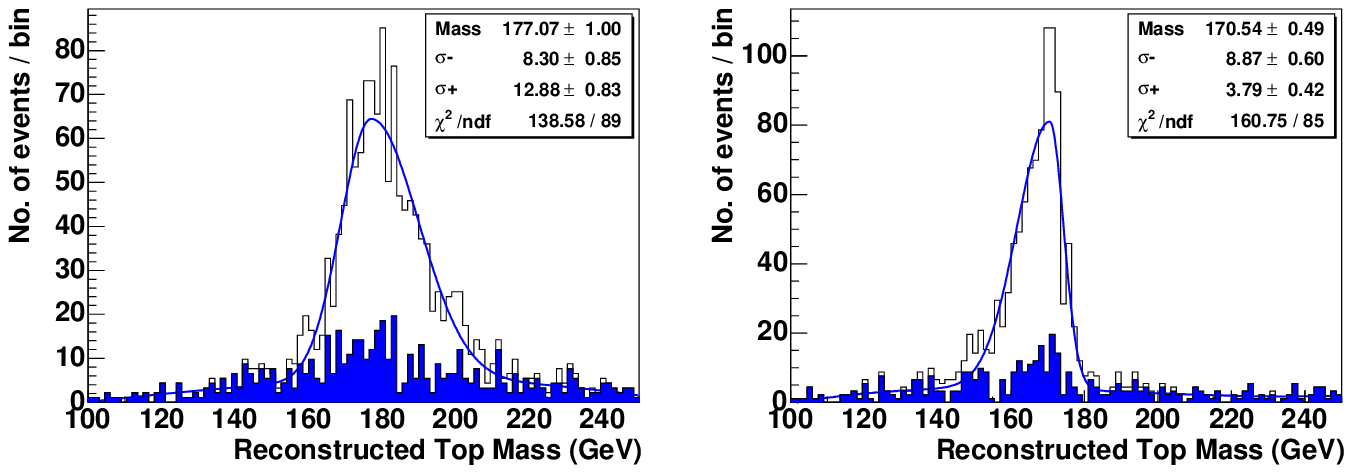}{The top mass distributions for (a) KtJet with a Fisher Cut value of 5.6 and $p_{t}cut$ of 45GeV; (b) PxCone with a cone Radius of 0.4 and $p_{t}cut$ of 45GeV. The contribution from combinatorial background is shown in blue.\label{top-mass}}

\subsection{Jet Algorithm Optimisation}
\label{opt}

In Sections~\ref{W-alg}~and~\ref{top-alg}, we presented results for a Fisher discriminant cut value of 5.6, and a cone radius of 0.4. In this section we shall justify these choices, and show the dependence of the purities and final efficiency on them.

In Section~\ref{jetalg} we motivated the study of a Fisher discriminant cut: to select both events which were `4-jet or 5-jet like'. Figure~\ref{fisher-purities} shows the dependence of the $W$ purity, $b$ purity, top purity and total efficiency on the choice of Fisher discriminant cut value for a $p_tcut$ value of 45GeV.
\EPSFIGURE[ht]{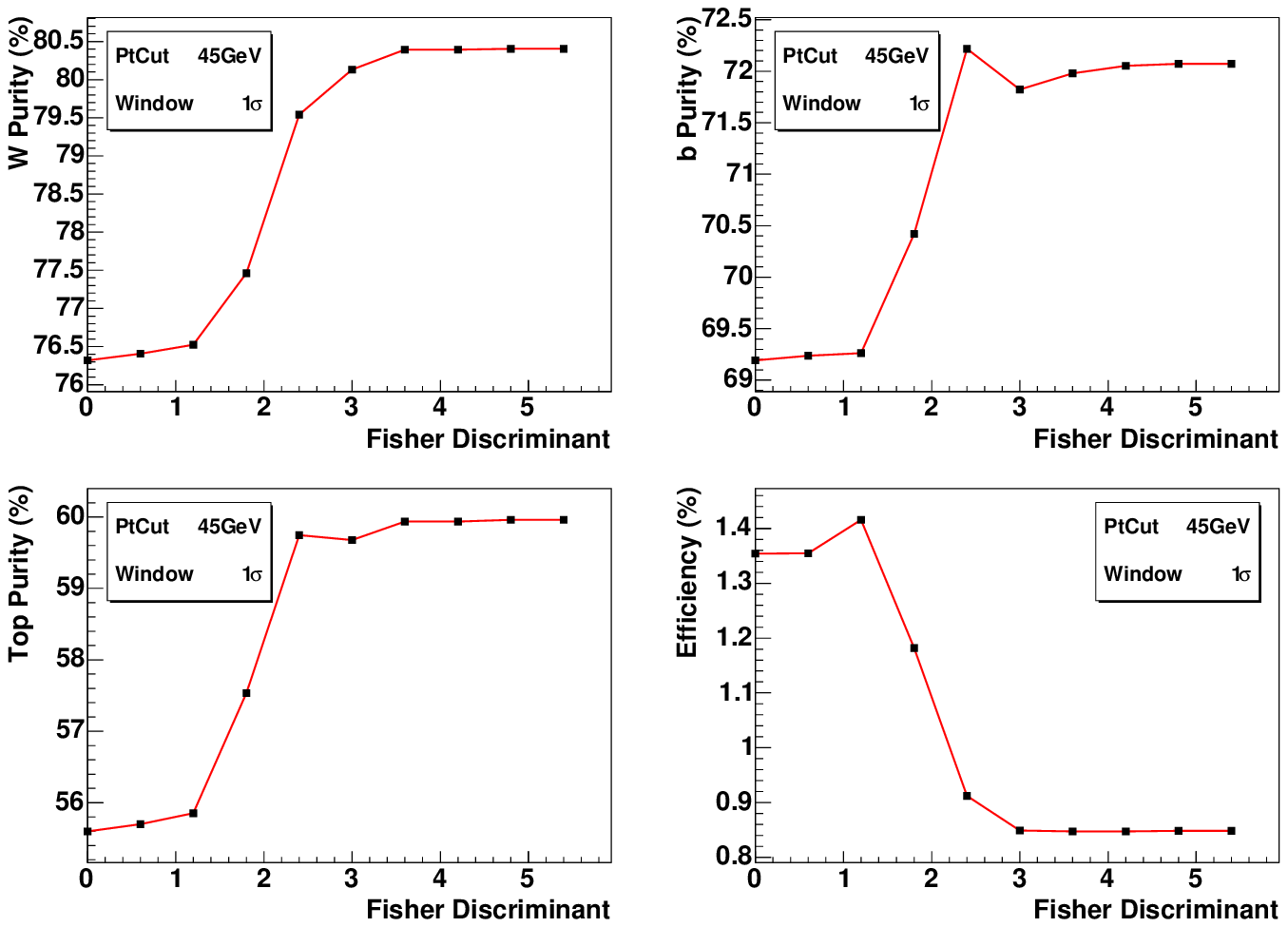}{The purities of (a) the hadronically decaying $W$ boson, (b) the $b$ jet, and (c) the top candidate, and (d) the total efficiency as a function of the Fisher discriminant cut for a $p_tcut$ value of 45GeV.\label{fisher-purities}}
We found that although the number of well reconstructed tops increases, such that the efficiency increases by almost a factor of 2 as the Fisher discriminant was varied between 0 (equivalent to clustering every event to 5 jets) and 5.4 (equivalent to clustering almost every event to 4 jets), the number of incorrect top candidates also increases by a similar amount, so that the purities of the reconstructed objects all decrease slightly. Since, as mentioned above, the LHC is expected to be a top factory, we have chosen a Fisher cut value of 5.4 to maximise the purity.

In the cone analysis, the dependence of these purities and efficiencies on the cone radius was studied. Figure~\ref{radius-purities} shows the dependence of the $W$ purity, $b$ purity, top purity and total efficiency respectively on the cone radius for a $p_tcut$ value of 45GeV.
\EPSFIGURE[ht]{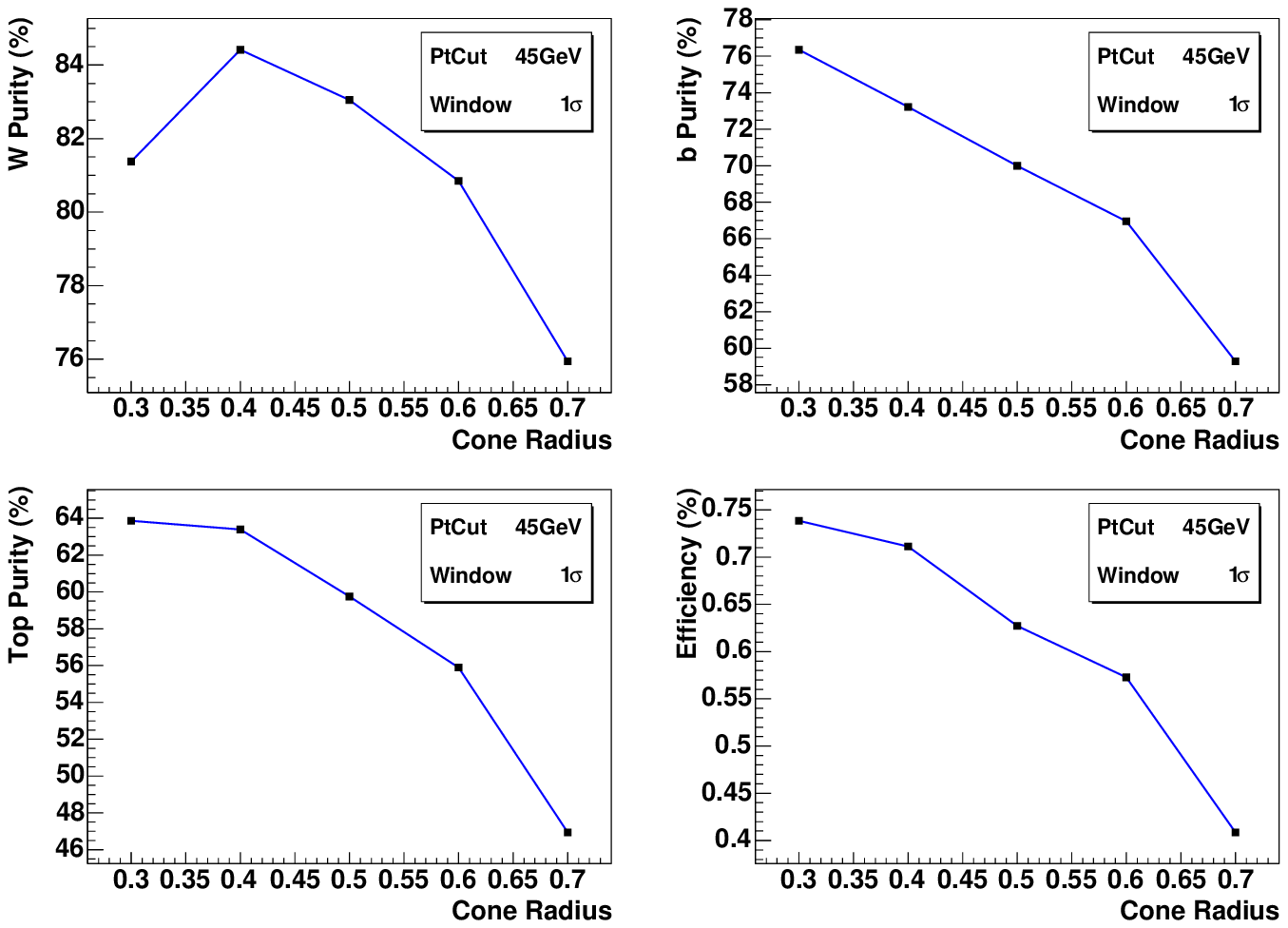}{The purities of (a) the hadronically decaying $W$ boson, (b) the $b$ jet, and (c) the top candidate, and (d) the total efficiency as a function of the cone radius for a $p_tcut$ value of 45GeV.\label{radius-purities}}
The cone radius that maximises the hadronic $W$ purity is quite different to the one that optimises the $b$ jet purity. Whereas the optimal cone radius for correctly reconstructing the $W$ seems to be 0.4 or 0.5, the optimal cone radius for reconstructing the $b$ jet seems to be 0.3 or less. Although the optimal cone radius for reconstructing the top quark seems to be 0.3, there is little difference in top purity between radii of 0.3 and 0.4. Since the larger radius seems to better reconstruct the $W$ boson, this was the radius chosen in this study.
\section{Background Study}
\label{background}

In this section we shall comment on the background processes $W+b\bar{b}+jets$ and $W+jets$, but we have not considered any other sources of background, for example production and decay of possible new supersymmetric particles, which in some models can give very large backgrounds to top production. The cross sections for the signal (inclusive $t\bar{t}$ production in the lepton plus jets channel), and these two background processes are shown in Table~\ref{proc}. The cross section for $W+b\bar{b}+jets$ is significantly smaller than that for inclusive $t\bar{t}$ production, therefore we expect the number of these events to be very small. We shall not consider it further. The cross sections for $W+n$jets for different multiplicities are shown in Table~\ref{wjets}.
\begin{table}[ht]
\begin{center}
\begin{tabular}{|l|r|}\hline
Process & Cross Section (pb) \\ \hline\hline
$t\bar{t}\rightarrow b\bar{b}\ell\nu_{\ell}jj$ & 235 \\ \hline
$\ell\nu_{\ell}+jets$ & 40200 \\ \hline
$\ell\nu_{\ell}+b\bar{b}+jets$ & 11.7 \\ \hline
\end{tabular}
\caption{The cross sections for signal and background processes, where $\ell=e,\mu$.}
\label{proc}
\end{center}
\end{table}
\begin{table}[ht]
\begin{center}
\begin{tabular}{|l|r|}\hline
Multiplicity & Cross Section (pb) \\ \hline\hline
n=0 & 33200 \\ \hline
n=1 & 4960 \\ \hline
n=2 & 1374 \\ \hline
n=3 & 670 \\ \hline
Total & 40204 \\ \hline
\end{tabular}
\caption{The cross sections for $W+n$jets as given by ALPGEN. We have used the default cuts and scale choice to generate the events, and imposed the MLM matching algorithm with the default parameters ($p_{T}^{jet}>20$GeV, $\Delta R_{jj}>0.7$).}
\label{wjets}
\end{center}
\end{table}

Since this is a generator level analysis, we have so far assummed a perfect $b$-tagging algorithm, i.e.~we assumed that 100\% of $b$-jets are so tagged, and 0\% of light jets are tagged. Therefore, if we were to apply the analysis algorithm described above to a sample of $W+jets$ events, one would expect almost\footnote{It is possible for the parton shower in HERWIG to give rise to a perturbative $b\bar{b}$ pair resulting in one or more jets. Our idealised $b$-tagging algorithm would tag such jets as $b$-jets.} all of these events to be removed, leaving a very large signal to background ratio. However, since the inclusive cross section for $W+jets$ at the LHC is expected to be large (two orders of magnitude larger than $t\bar{t}$ production), by allowing the possibility of  mis-tagging light jets as $b$-jets one might still expect a significant number of these events to pass the selection cuts. In addition to this, in a realistic experimental environment not all $b$-jets are so tagged, so we also allowed for this in order to reduce the number of signal events expected to a more realistic level

Therefore we applied a $b$-tagging efficiency to each jet in the signal and $W+jets$ samples. These efficiencies were taken as 60\% for $b$-jets and 1\% for light jets. The final cross sections for the signal and background processes after all cuts are shown in Table~\ref{fcs}. We conclude that at this level the backgrounds are negligible. 

Naively one might expect that by clustering every event to four jets, the KtJet analysis will have more background events than the cone algorithm analysis. This is because events from the lower multiplicity samples may pass the selection cuts, and since these lower multiplicity samples have a large cross section even if a small fraction of them pass the cuts this could lead to a significant increase in background levels. However, we do not find this to be a problem as no events from the lower multiplicity samples ($n\leq 2$) pass the final selection cuts in the KtJet analysis\footnote{Each of these samples contained $\sim 10^{6}$ events.}.
\begin{table}[ht]
\begin{center}
\begin{tabular}{|l|r|r|}\hline
 & \multicolumn{2}{c|}{Final Cross Section (pb)} \\ 
\cline{2-3}
Process & KtJet & PxCone \\ \hline\hline
$t\bar{t}\rightarrow b\bar{b}\ell\nu_{\ell}jj$ & 3.04 & 2.45 \\ \hline
$W$+0jets & $<0.16$ & $<0.16$ \\ \hline
$W$+1jet & $<1\times10^{-2}$ & $<1\times10^{-2}$ \\ \hline
$W$+2jets & $<5\times10^{-3}$ & $<5\times10^{-3}$ \\ \hline
$W$+3jets & $\sim 3\times10^{-3}$ & $\sim 2\times10^{-3}$ \\ \hline
\end{tabular}
\caption{The final cross sections for signal and background processes. In cases where no events passed the selection cuts, we quote upper limits at the 95\% confidence level.}
\label{fcs}
\end{center}
\end{table}

\section{Systematic Errors}
\label{syst}

The inclusive $t\bar{t}$ cross section at the LHC will be very high, and as a result the statistical error on the top mass measurement is expected to be small. Therefore the systematic error will be the dominant source of uncertainty, and in order to make a high precision measurement, a good understanding of this will be required. In this section, we shall discuss some of the sources of systematic errors.

\subsection{Sources of systematics}

There are several sources of systematic errors on the top mass measurement in the lepton plus jets channel. These include effects that are present at the generator level, as well as detector effects such as the Jet Energy Scale; since we have not applied any detector simulation here we shall only discuss the former. The two main effects we shall consider are Initial- and Final-State Radiation (ISR/FSR); and the Underlying Event (UE). We shall also comment on the contribution to the systematic error from hadronisation effects. 

It is a well know result of QCD that both the initial-state and final-state partons involved in a scattering process will emit additional partons. Clearly it is neither of these effects alone that is a good (gauge-invariant) observable but their sum, i.e.~all partons emitted from the scattering partons as a whole. However, these two types of emission will produce different systematic shifts on the top mass - while FSR\footnote{By FSR we primarily mean FSR from the decay products of the top. FSR from the tops themselves is relatively rare since the cross section is dominated by the region where they are slow moving.} could cause radiation out of the three jets resulting in a lower mass peak, ISR could cause radiation into the jets resulting in a higher mass peak. Thus, if one were to consider these effects together to estimate the systematic error on the top mass, one could underestimate the error. For this reason, the two effects are often considered separately, and this is what we shall do here.

At the LHC the cross section for multiple interactions is expected to be high. The presence of additional hadronic activity from such scatters could cause a shift of the top peak to higher mass values. As this is a non-perturbative effect, one must use a phenomenological model, parameterising the underlying event in terms of a finite number of tuneable parameters that must be fit to data. There are several competing models (eg JIMMY, PYTHIA, SHERPA) and these all predict different UE multiplicities at the LHC. One could therefore use this difference as an estimate of the UE systematic, however, we would expect that once LHC data becomes available one would be able to rule out some of these models in favour of others. So a more realistic handle on the UE systematic may be to vary some of the parameters of a particular model within accepted values.

\subsection{Dominant Systematic Errors}

One might expect these systematic errors to be quite different for the two jet algorithms considered here. While one would expect the cone algorithm measurement with a small cone radius to be dominated by FSR and little affected by ISR and UE effects; one would expect that the KtJet measurement would suffer less from FSR, but more from ISR and UE. In order to illustrate this, we have run different parton level outputs through our analysis code. 
\EPSFIGURE[ht]{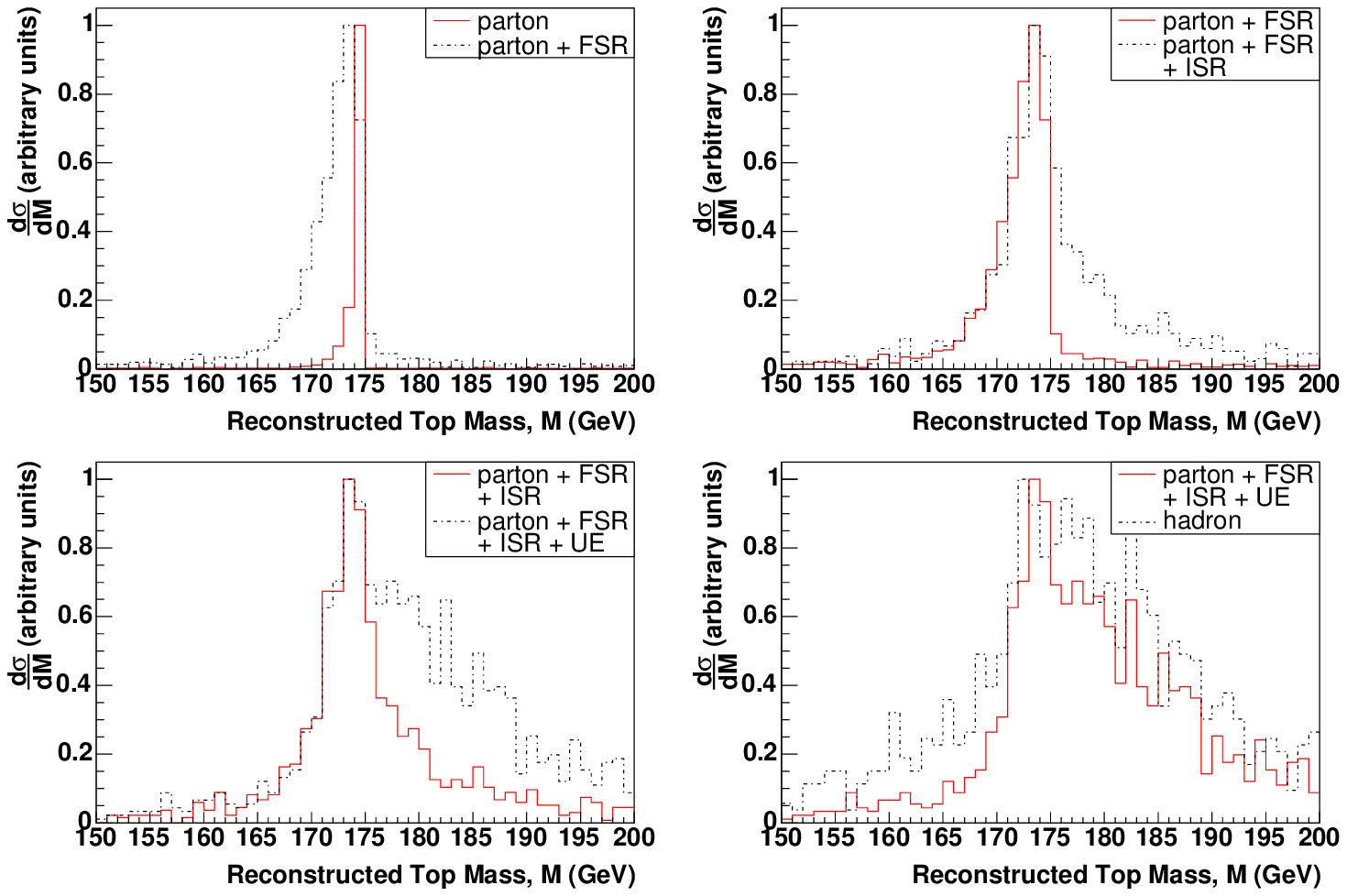}{The effects on the top mass peak reconstructed by KtJet when adding in (a) partons from Final State Radiation; (b) partons from Initial State Radiation; (c) partons from multiple scattering and (d) Hadronisation effects.\label{syst-toy-kt}}
\EPSFIGURE[ht]{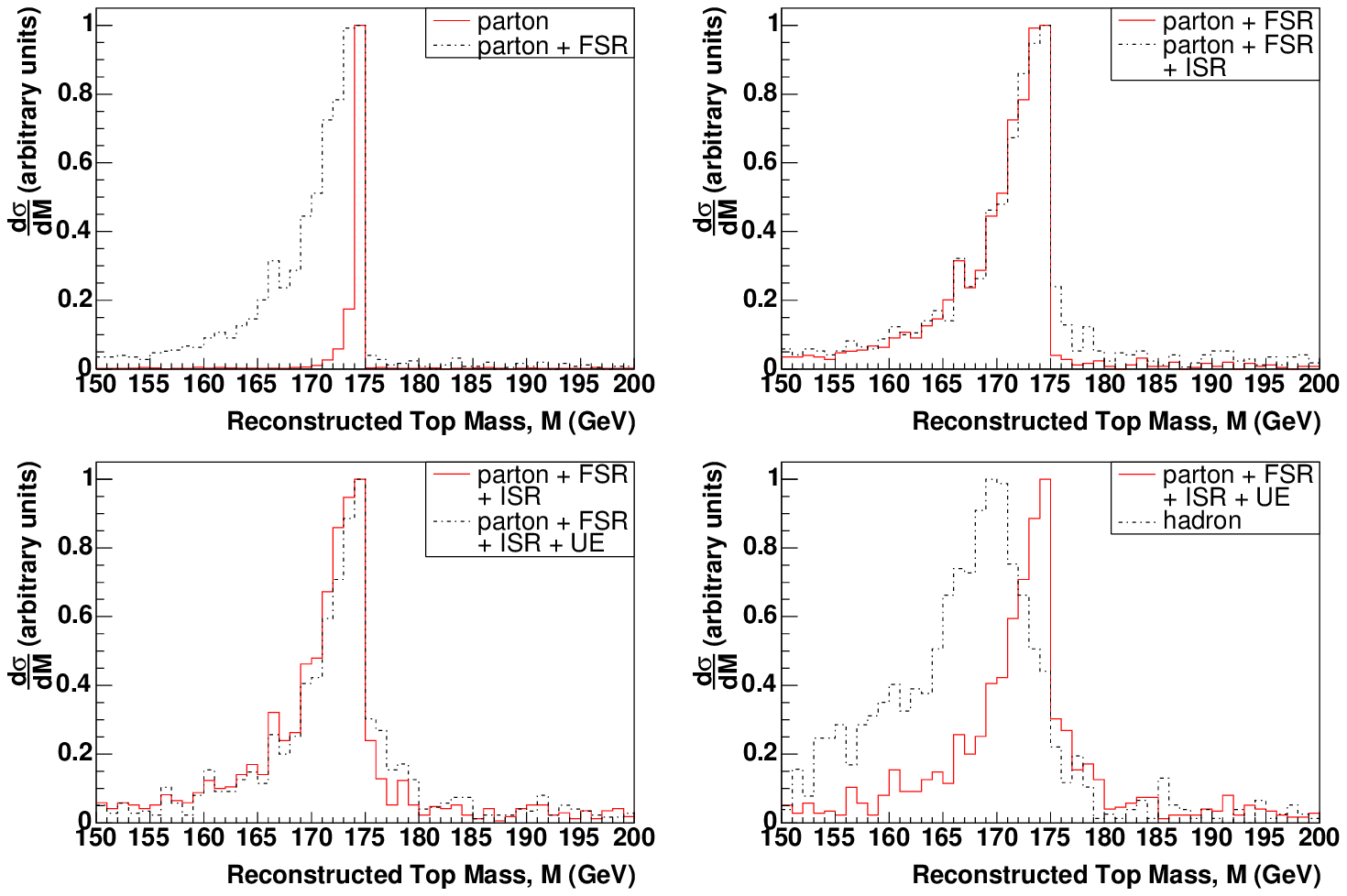}{The effects on the top mass peak reconstructed by PxCone when adding in (a) partons from Final State Radiation; (b) partons from Initial State Radiation; (c) partons from multiple scattering and (d) Hadronisation effects.\label{syst-toy-cone}}

The different parton level outputs selected were: the partons from the hard subprocess; the partons from FSR; the partons from both FSR and ISR, and the partons from FSR, ISR and UE. Thus one can see these different effects causing the top mass to shift and broaden out in several stages. These results are shown in Figures~\ref{syst-toy-kt}~and~\ref{syst-toy-cone}, where the hadron level peaks are also shown to illustrate the effect of hadronisation.

These figures do indicate that these different effects which contribute to the systematic error do so differently for the two jet algorithms. We do find that the cone algorithm systematic is dominated by the FSR, and that it is little affected by the ISR and the UE. In contrast the three effects seem to contribute equally to the KtJet systematic. This may be important when data becomes available, and the LHC experiments start to more accurately estimate this systematic error. We also note that hadronisation also affects the mass reconstructed by both algorithms. Both reconstructed top peaks are broadened by these effects; and the peak position is shifted significantly in the case of the cone algorithm. Since hadronisation is non-perturbative physics, and so not very well understood, having two algorithms which are affected differently by hadronisation may help to estimate this contribution to the systematic error. We conclude that the two algorithms have very different contributions to their systematic errors and therefore that a study combining information from both would be more precise than either alone.

\section{Conclusion}

In this analysis, we have compared the abilities of two jet algorithms to reconstruct the top mass in the lepton plus jets channel at the LHC. The two jet algorithms used were a traditional mid-point iterating cone algorithm, and the $k_{\perp}$ clustering algorithm, KtJet. Each of the jet algorithms was optimised separately.

KtJet was run in the exclusive mode, and the events were clustered until there were $N$ jets, where $N$ was either 4 or 5. KtJet contains much of the structure of the event in the perturbative merging scales, $d_{i+1,i}^{1/2}$, and a linear combination of these was used to choose $N$. The linear combination chosen was the `Fisher Discriminant', $\mathscr{F}$. This was calculated for each event, and if $\mathscr{F}<\mathscr{F}_{cut}$, then the event was clustered to 4 jets, otherwise it was clustered to 5 jets. KtJet was optimised by finding the value of $\mathscr{F}_{cut}$ which maximised the purity of the reconstructed tops. It was found that the optimum value was one which corresponded to always clustering to 4 jets, although by varying $\mathscr{F}_{cut}$, such that some events were clustered to 5 jets resulted in an increase in efficiency almost by a factor of 2.

The cone algorithm has a radius parameter, and was optimised by finding the radius that maximised the purity of the reconstructed tops, and efficiency with which they were reconstructed. We found this to be 0.4.

We found that both algorithms when optimised yielded similar values for the purity of the reconstructed objects, and the final efficiency. However, the mass resolution was better when using the cone algorithm.

We considered the processes $W+b\bar{b}+jets$ and $W+jets$ as possible sources of background at the LHC. The cross section for $W+b\bar{b}+jets$ was found to be significantly smaller than that for $t\bar{t}$ production. We applied a slightly more realistic $b$-tagging algorithm to the signal and $W+jets$ samples such that only 60\% of all $b$-jets, and 1\% of all light jets were tagged. We found that even with this more realistic algorithm, the level of background passing the selection cuts of our analysis was negligible.

Finally, we discussed some of the contributions to the systematic error on the top mass in this channel that are present at the generator level. The sources of systematic error considered were final state radiation (FSR), initial state radiation (ISR), the underlying event (UE) and hadronisation. Using parton level Monte Carlo truth, we were able to isolate these effects, and study them individually. 

We found that in the cone algorithm analysis the systematic error was dominated by FSR, energy was radiated out of the cones resulting in a lower reconstructed top mass, and the other two processes had little effect on the result. In the KtJet analysis, we found that the contribution to the systematic error from FSR was smaller, but the contributions from ISR and the UE larger. We also found that the cone algorithm was more affected by hadronisation effects than KtJet. 

The fact that the sources of systematic error on the top mass are very different for these two algorithms is very encouraging for the top mass measurement at the LHC. Since the inclusive $t\bar{t}$ cross section will be very large at the LHC, the statistical error is expected to be small. In order to make a precise measurement of the top mass, one will need to understand the different effects that contribute to the systematic error very well to reduce the error on the top mass. It is worth pointing out that the corrections from FSR and ISR are perturbative effects and so the corrections from these could, at least in principle, be quantified. However corrections from the underlying event and hadronisation are fundamentally non-perturbative. The fact that KtJet suffers mainly from the underlying event and the cone algorithm suffers mainly from hadronisation again means that combining information from both may help to better understand these non-perturbative corrections. Since these two jet algorithms have such complementary systematic errors, by using both on the data one might hope to better constrain the systematic error on the top mass and so make a more precise measurement.

\section*{Acknowledgments}

We would like to thank Roger Barlow, George Lafferty, Fred Loebinger, Andrew Pilkington and Thorsten Wengler for many useful and interesting discussions and suggestions. We would also like to thank the U.K. Particle Physics and Astronomy Research Council for funding this work.


\end{document}